\documentclass[10pt, a4paper]{amsart}

\usepackage{amsmath, amsfonts, amssymb, mathtools, url, graphicx, graphicx}
\usepackage{newtxtext}
\usepackage{newtxmath}

\usepackage{enumitem}
\usepackage[usenames, dvipsnames]{color}
\allowdisplaybreaks
\usepackage{tikz}
\usetikzlibrary{automata,positioning}
\usepackage{caption}
\usepackage{subcaption}
\usepackage{algorithm,algorithmic}
\usepackage{todonotes}
\usepackage{empheq}
\newtheorem{lemma}{Lemma}
\newtheorem{proposition}{Proposition}
\newtheorem{defn}{Definition}
\newtheorem{assump}{Assumption}

\newtheorem{prob}{Problem}
\newtheorem{remark}{Remark}

\newtheorem{fact}{Fact}

\newcommand{\norm}[1]{\left\lVert{#1}\right\rVert}
\newcommand{\abs}[1]{\left\lvert{#1}\right\rvert}

\newcommand{\pmat}[1]{\begin{pmatrix}#1\end{pmatrix}}

\newcommand{\R}{\mathbb{R}}
\newcommand{\N}{\mathbb{N}}
\renewcommand{\P}{\mathcal{P}}
\newcommand{\PS}{\P_{S}}
\newcommand{\PU}{\P_{U}}
\newcommand{\Sw}{\mathcal{S}}
\newcommand{\wv}{\overline{w}}
\newcommand{\we}{\underline{w}}

\DeclareMathOperator{\minimize}{minimize}
\DeclareMathOperator{\sbjto}{subject\;to}

\allowdisplaybreaks

\title[]{Data-driven switching logic design for switched linear systems}

\author{Atreyee Kundu}
\thanks{The author is with the Department of Electrical Engineering, Indian Institute of Science Bangalore, India. Email: atreyeek@iisc.ac.in.}
\keywords{switched linear systems, data-based design, stabilizing switching logics}
\date{\today}

\begin{document}

\maketitle

	\begin{abstract}
        This paper deals with stabilization of discrete-time switched linear systems when explicit knowledge of the state-space models of their subsystems is not available. Given the set of admissible switches between the subsystems, the admissible dwell times on the subsystems and a set of finite traces of state trajectories of the subsystems that satisfies certain properties, we devise an algorithm that designs periodic switching logics which preserve stability of the resulting switched system. We combine two ingredients: (a) data-based stability analysis of discrete-time linear systems and (b) multiple Lyapunov-like functions and graph walks based design of stabilizing switching logics, for this purpose. A numerical example is presented to demonstrate the proposed algorithm.
	\end{abstract}

\section{Introduction}
\label{s:intro}
\subsection{Motivation}
\label{ss:motive}
    Hybrid systems find wide applications in modern day Cyber-Physical Systems (CPS). In this paper we deal with a subclass of hybrid systems, where we focus on the discrete dynamics and abstract away the continuous dynamics as switching. Such a system is known as discrete-time \emph{switched system} and contains two components --- a family of systems and a switching logic. The \emph{switching logic} selects an \emph{active subsystem} at every instant of time, i.e., the system from the family whose dynamics is currently being followed \cite[\S 1.1.2]{Liberzon}.

    It is well-known that a switched system does not necessarily inherit qualitative properties of its constituent subsystems. For instance, divergent trajectories may be generated by switching appropriately among stable subsystems while a suitably constrained switching logic may ensure stability of a switched system even if all subsystems are unstable (see e.g., \cite[p.\ 19]{Liberzon} for examples with two subsystems). Due to such interesting features, the problem of designing stabilizing switching logics for switched systems has attracted considerable research attention in the past few decades, see e.g., the survey papers \cite{Liberzon_survey, Heemels_survey, Antsaklis_survey} and the references therein.

    The said design problem typically requires the availability of mathematical models of the subsystems. However, on the one hand, in many real-world scenarios, particularly for large-scale complex systems, accurate mathematical models of the subsystems, such as transfer functions, state-space models or kernel representations, are often difficult to infer. On the other hand, we often have access to traces of state trajectories of the subsystems. Such trajectories may be obtained from a simulation model provided by the system manufacturer to study the system behaviour with respect to various sets of inputs prior to their application to the actual system, or from an experiment conducted during the operations of the subsystems. The above interesting facts motivate our research on data-driven design of switching logics that ensure good qualitative and/or quantitative properties of switched systems. In this paper we present an algorithm to design stabilizing switching logics for discrete-time switched linear systems when explicit knowledge of the state-space models of their subsystems are not available. We restrict our attention to sets of switching logics that obey pre-specified restrictions on admissible switches between the subsystems and admissible dwell times on the subsystems.
\subsection{Literature survey}
\label{ss:lit_survey}
    Stability analysis and control synthesis of switched systems without explicitly involving mathematical models of their subsystems, are dealt with recently in \cite{Kenarian2019,Li2018,def}. The work \cite{Kenarian2019} addresses the problem of deciding stability of a discrete-time switched linear system under arbitrary switching logics from a set of finite traces of state trajectories. Probabilistic stability guarantees are provided as a function of the number of available state observations and a desired level of confidence. In \cite{Li2018} reinforcement learning techniques are employed for optimal control of switched linear systems. A Q-learning based algorithm is proposed to design a discrete switching logic and a continuous control logic such that a certain infinite-horizon cost function is minimized. The convergence guarantee of the proposed algorithm is, however, not available. A randomized polynomial-time algorithm for the design of switching logics under the availability of certain information about the multiple Lyapunov(-like) functions \cite[\S3.1]{Liberzon} corresponding to the individual subsystems in an expected sense, is presented in \cite{def}. The authors show that if it is allowed to switch from any subsystem to a certain number of stable subsystems, then a switching logic obtained from the proposed algorithm is stabilizing with overwhelming probability.
\subsection{Our contributions}
\label{ss:contri}
    We assume that the knowledge of the set of admissible switches between the subsystems and the admissible dwell times on the subsystems is available. In addition, we have access to a set of finite traces of state trajectories of the subsystems that satisfies certain properties (henceforth, to be referred to as \emph{subsystems data}). We combine two ingredients: (a) data-based techniques for stability analysis of discrete-time linear systems and (b) multiple Lyapunov-like functions and graph walks based techniques for the design of stabilizing switching logics for switched systems, towards arriving at an algorithm that designs stabilizing periodic switching logics in the absence of accurate state-space models of the subsystems. Our algorithm involves the following steps:
    \begin{itemize}[label = \(\circ\), leftmargin = *]
        \item First, we determine (in)stability of the subsystems based on the subsystems data, and compute multiple Lyapunov-like functions and a set of corresponding scalars.
        \item Second, a class of cycles on the underlying directed graph that satisfies certain conditions involving the above set of scalars and the admissible dwell times on the subsystems, is identified.
        \item Third, a stabilizing switching logic is designed such that it activates the vertices that appear in the above cycles and dwells on them for appropriate durations. If no favourable cycle is found on the underlying directed graph of the switched system, then the algorithm reports a failure.
    \end{itemize}

     The design of stabilizing switching logics for switched systems by involving multiple Lyapunov-like functions and graph-theoretic tools is standard in the literature, see e.g., \cite{abc,xyz3,xyz4} and the references therein. However, the suitable Lyapunov-like functions employed in this task, are commonly designed under complete knowledge of the state-space models of the subsystems. We assume certain properties of the available dataset, and involve techniques of data-based stability analysis of linear systems proposed in \cite{Park2009} to design multiple Lyapunov-like functions and compute a set of relevant scalars. These scalars together with the admissible switches between the subsystems and admissible dwell times on the subsystems are then employed to design stabilizing switching logics. At this point, it is worth highlighting that we do not opt for the construction of state-space models of the subsystems from the given data, and hence the proposed technique does not involve system identification of the switched system under consideration. To the best of our knowledge, this is the first instance in the literature where the design of stabilizing switching logics for switched systems in the setting of multiple Lyapunov-like functions and graph-theoretic tools is addressed without explicit knowledge of the state-space models of the subsystems.
\subsection{Paper organization}
\label{ss:paper_org}
    The remainder of this paper is organized as follows: In \S\ref{s:prob_stat} we formulate the problem under consideration. A set of preliminaries required for our result is presented in \S\ref{s:prelims}. Our result appears in \S\ref{s:mainres}. We present a numerical example in \S\ref{s:numex}, and conclude in \S\ref{s:concln} with a brief discussion of future research directions.
\subsection{Notation}
\label{ss:notation}
    \(\R\) is the set of real numbers and \(\N\) is the set of natural numbers, \(\N_{0} = \N\cup\{0\}\). We denote by \([k_{1}:k_{2}]\) the set \(\{n\in\N_{0}\::\:k_{1}\leq n\leq k_{2}\}\). \(I_{d}\) denotes the \(d\times d\) identity matrix, \(I_{d}^{k}\) denotes its \(k\)-th column, \(k\in\{1,2,\ldots,d\}\), \(0_{n}\) and \(\overline{0}_{n}\) denote the \(d\times 1\) zero matrix and \(d\times d\) zero matrix, respectively. For a matrix \(B\in\R^{d\times d}\), \(B\succ 0\) (resp., \(B\prec 0\)) denotes that \(B\) is positive definite (resp., negative definite), and \(\lambda_{\max}(B)\) denotes the maximal eigenvalue of \(B\).
\section{Problem statement}
\label{s:prob_stat}
    We consider a family of discrete-time linear systems
    \begin{align}
    \label{e:family}
        x(t+1) = A_{i}x(t),\:\:x(0) = x_{0},\:\:i\in\P,\:\:t\in\N_{0},
    \end{align}
    where \(x(t)\in\R^{d}\) is the vector of states at time \(t\), \(\P = \{1,2,\ldots,N\}\) is an index set, and
    \begin{align}
    \label{e:comp_form}
        A_{i} = \pmat{-a_{i,d-1}&\cdots&-a_{i,1}&-a_{i,0}\\1&\cdots&&0\\&\vdots&&\vdots\\&\cdots&1&0}\in\R^{d\times d},\:\:i\in\P
    \end{align}
    are full-rank constant matrices. Let \(\sigma:\N_{0}\to\P\) be a switching logic. A discrete-time switched linear system generated by the family of systems \eqref{e:family} and a switching logic \(\sigma\) is described as
    \begin{align}
    \label{e:swsys}
        x(t+1) = A_{\sigma(t)}x(t),\:\:x(0) = x_{0},\:\:t\in\N_{0},
    \end{align}
    where we have suppressed the dependence of \(x\) on \(\sigma\) for notational simplicity.

    Let \(\PS\) and \(\PU\) denote the sets of indices of the Schur stable and unstable subsystems, respectively, \(\P = \PS\sqcup\PU\).\footnote{A matrix \(B\in\R^{d\times d}\) is Schur stable if all its eigenvalues are inside the open unit disk. We call \(B\) unstable if it is not Schur stable.} We let \(E(\P)\) be the set of ordered pairs \((i,j)\) such that a switch from subsystem \(i\) to subsystem \(j\) is admissible, \(i,j\in\P\), \(i\neq j\), and \(\delta\in\N\) and \(\Delta\in\N\) be the admissible minimum and maximum dwell times on all subsystems, respectively, \(\delta\leq\Delta\). Let \(0=:\tau_{0}<\tau_{1}<\tau_{2}<\cdots\) be the \emph{switching instants}; these are the points in time where \(\sigma\) jumps. A switching logic \(\sigma\) is called \emph{admissible} if it satisfies the following conditions: \((\sigma(\tau_{k}),\sigma(\tau_{k+1}))\in E(\P)\) and \(\tau_{k+1}-\tau_{k}\in[\delta:\Delta]\), \(k=0,1,2,\ldots\). Let \(\Sw\) denote the set of all admissible \(\sigma\). We are interested in stability of the switched system \eqref{e:swsys}. Recall that
    \begin{defn}
    The switched system \eqref{e:swsys} is \emph{globally asymptotically stable} (GAS) for a given switching logic $\sigma$ if \eqref{e:swsys} is Lyapunov stable and globally asymptotically convergent, i.e., for all $x(0)$, $\displaystyle{\lim_{t\rightarrow+\infty}\norm{x(t)} = 0}$.
    \end{defn}

    The scalars \(a_{i,j}\), \(j=0,1,\ldots,d-1\), \(i\in\P\), are unknown. The set \(E(\P)\) and the scalars \(\delta\), \(\Delta\) are known. In addition, finite traces of state trajectories \(x(0), x(1),\ldots,x(L)\), \(L\in\N\), are available for all subsystems. We will solve the following problem:
    \begin{prob}
    \label{prob:mainprob}
    \rm{
        Given the set of admissible switches between the subsystems, \(E(\P)\), the admissible minimum and maximum dwell times on the subsystems, \(\delta\) and \(\Delta\), and a subsystems dataset, design a switching logic \(\sigma\) that preserves GAS of the switched system \eqref{e:swsys}.
        }
    \end{prob}

    \begin{remark}
    \label{rem:knowledge}
    \rm{
        We observe that availability of the knowledge of the set of admissible switches between the subsystems and the dwell times on the subsystems is no loss of generality. Indeed, admissible switches between the subsystems are often governed by properties of internal components of a system. A distinction between admissible and inadmissible transitions captures situations where switches between certain subsystems may be prohibited. For instance, in automobile gear switching, certain switching orders (e.g., from first gear to second gear, etc.) are followed \cite[\S III]{Antsaklis_survey}. A restriction on minimum dwell time on subsystems arises in situations where actuator saturations may prevent switching frequency beyond a certain limit. Also, in order to switch from one component to another, a system may undergo certain operations of non-negligible durations \cite[\S I]{Heydari_ACC16} resulting in a minimum dwell time restriction on subsystems. Restricted maximum dwell time is natural to systems whose components need regular maintenance or replacements, e.g., aircraft carriers, MEMS systems, etc. Moreover, systems dependent on diurnal or seasonal changes, e.g., components of an electricity grid, have inherent restrictions on admissible dwell time. As a result, the set \(E(\P)\) and the scalars \(\delta\) and \(\Delta\) are naturally pre-specified and not design parameters.
        }
    \end{remark}

        Notice that even though we are allowing stable subsystems, Problem \ref{prob:mainprob} does not admit a trivial solution. Indeed, under the maximum dwell time restriction, finding a stable subsystem \(i\in\P_S\) by employing the available subsystems data and setting \(\sigma(t)=i\) for all \(t\in\N_{0}\) is not allowed. Prior to presenting our solution to Problem \ref{prob:mainprob}, we catalog a set of preliminaries.
\section{Preliminaries}
\label{s:prelims}
\subsection{Multiple Lyapunov-like functions and graph walks based design of stabilizing switching logics}
\label{ss:swsig_design}
    The following fact is well-known:
    \begin{fact}{\cite[Fact 1]{abc}}
    \label{fact:key1}
    \rm{
        For each \(i\in\P\), there exists a pair \((P_{i},\lambda_{i})\in\R^{d\times d}\times\R\), where \(P_{i}\) is a symmetric and positive definite matrix, and \(0 < \lambda_{i} < 1\), if \(i\in\PS\) and \(\lambda_{i}\geq 1\), if \(i\in\PU\), such that, with
        \begin{align}
        \label{e:Lyap_eq1}
            \R^{d}\ni\xi\mapsto V_{i}(\xi) := \xi^\top P_{i}\xi\in[0,+\infty[,
        \end{align}
        we have
        \begin{align}
        \label{e:Lyap_eq2}
            V_{i}(\gamma_{i}(t+1))\leq\lambda_{i}V_{i}(\gamma_{i}(t)),\:\:t\in\N_{0},
        \end{align}
        and \(\gamma_{i}(\cdot)\) solves the \(i\)-th recursion in \eqref{e:family}.
        }
    \end{fact}
    The functions \(V_{i}\), \(i\in\P\) are Lyapunov-like functions corresponding to the subsystems \(i\in\P\). The scalar \(\lambda_{i}\), \(i\in\P\) gives a quantitative measure of (in)stability of subsystem \(i\). The Lyapunov-like functions corresponding to the individual subsystems are related as follows:
    \begin{fact}{\cite[Fact 2]{abc}}
    \label{fact:key2}
    \rm{
        There exists \(\R\ni\mu_{ij}>0\) such that
        \begin{align}
        \label{e:Lyap_eq3}
            V_{j}(\xi)\leq\mu_{ij}V_{i}(\xi)\:\:\text{for all}\:\xi\in\R^{d},\:(i,j)\in E(\P).
        \end{align}
        }
    \end{fact}
    A \emph{tight} estimate of \(\mu_{ij}\), \((i,j)\in E(\P)\), is provided below.
    \begin{proposition}{\cite[Proposition 1]{abc}}
    \label{prop:mu_estimate}
    \rm{
        The scalars \(\mu_{ij}\), \((i,j)\in E(\P)\) can be computed as follows:
        \begin{align}
        \label{e:mu_estimate}
            \mu_{ij} = \lambda_{\max}(P_{j}P_{i}^{-1}),\:\:(i,j)\in E(\P).
        \end{align}
        }
    \end{proposition}

    We associate a directed graph \(G(\P,E(\P))\) with the family of systems \eqref{e:family} and the set of admissible switches between the subsystems, \(E(\P)\), as follows: the set of vertices of \(G\) is the set of indices of the subsystems, \(\P\), and the set of edges of \(G\) is the set of admissible switches between the subsystems, \(E(\P)\). A \emph{walk} on \(G\) is a finite alternating sequence of vertices and edges, \(W = v_{0},(v_{0},v_{1}),v_{1},\ldots,v_{\ell-1},(v_{\ell-1},v_{\ell}),v_{ell}\), where \(v_{k}\in\P\), \((v_{k},v_{k+1})\in E(\P)\), \(k=0,1,\ldots,\ell-1\). \(W\) is a \emph{cycle} if \(v_{\ell}:= v_{0}\) and \(v_{0},v_{1},\ldots,v_{\ell-1}\) are distinct.

    Let \(\wv:\P\to\R\) and \(\we:E(\P)\to\R\) be defined as
    \begin{align*}
        \wv(i) =
        \begin{cases}
            -\abs{\ln\lambda_{i}},\:\:&\:\text{if}\:i\in\PS,\\
            \abs{\ln\lambda_{i}},\:\:&\:\text{if}\:i\in\PU,
        \end{cases}
    \end{align*}
    and
    \begin{align*}
        \we(i,j) = \ln\mu_{ij},\:\:&\:\:(i,j)\in E(\P),
    \end{align*}
    where the scalars \(\lambda_{i}\), \(i\in\P\) and \(\mu_{ij}\), \((i,j)\in E(\P)\) are as described in Facts \ref{fact:key1} and \ref{fact:key2}, respectively.
    \begin{defn}
    \label{d:contrac}
    \rm{
        A cycle \(W = v_{0},(v_{0},v_{1}),v_{1},\ldots,v_{\ell-1},(v_{\ell-1}\),\(v_{0}),v_{0}\) on \(G\) is called \emph{contractive} if there exist \(D_{v_{k}}\in[\delta:\Delta]\), \(k=0,1,\ldots,\ell-1\), such that the following condition holds:
        \begin{align}
        \label{e:contrac}
            \sum_{k=0}^{\ell-1}\wv(v_{k})D_{v_{k}} + \sum_{\substack{{k=0}\\{v_{\ell}:=v_{0}}}}^{\ell-1}\we(v_{k},v_{k+1}) < 0.
        \end{align}
        }
    \end{defn}

    Given a contractive cycle \(W = v_{0},(v_{0},v_{1}),v_{1},\ldots,v_{\ell-1}\),\\\((v_{\ell-1},v_{0}),v_{0}\) on \(G\) and the corresponding \(D_{v_{0}}\), \(D_{v_{1}},\ldots,D_{v_{\ell-1}}\), the following algorithm \cite[Algorithm 1]{xyz4} constructs a switching logic \(\sigma\).
    \begin{algorithm}[htbp]
			\caption{Construction of switching logics} \label{algo:swsig_construc}
		\begin{algorithmic}[1]
			\renewcommand{\algorithmicrequire}{\textbf{Input:}}
			\renewcommand{\algorithmicensure}{\textbf{Output:}}
			
			\REQUIRE A contractive cycle, \(W = v_{0},(v_{0},v_{1}),v_{1},\ldots,v_{\ell-1}\),\((v_{\ell-1},v_{0}),v_{0}\) on \(G\) and the corresponding \(D_{v_{0}},D_{v_{1}},\ldots,D_{v_{\ell-1}}\).
			\ENSURE A switching logic, \(\sigma\).
			
            \STATE Set \(p=0\) and \(\tau_{0} = 0\).
            \FOR {\(k=p\ell,p\ell+1,\ldots,(p+1)\ell-1\)}
                \STATE Set \(\sigma(\tau_{k}) = v_{k-p\ell}\) and \(\tau_{k+1} = \tau_{k} + D_{v_{k-p\ell}}\).
			 \ENDFOR
            \STATE Set \(p=p+1\) and go to 2.
		\end{algorithmic}
	\end{algorithm}

    By construction, \(\sigma\) is periodic with period \(\displaystyle{\sum_{k=0}^{\ell-1}D_{v_{k}}}\).
    \begin{lemma}{\cite[Theorem 1]{xyz4}}
    \label{lem:stab}
    \rm{
        Consider a switching logic \(\sigma\) obtained from Algorithm \ref{algo:swsig_construc}. The following are true:
        \begin{enumerate}[label = \roman*), leftmargin = *]
            \item \(\sigma\in\Sw\), and
            \item \eqref{e:swsys} is GAS under \(\sigma\).
        \end{enumerate}
        }
    \end{lemma}

    Algorithm \ref{algo:contrac_cycle_detection} \cite[Proposition 2]{xyz3} is geared towards detecting a contractive cycle on \(G\).
    \begin{algorithm}[htbp]
	\caption{Detection of contractive cycle on \(G\)} \label{algo:contrac_cycle_detection}
    	\begin{algorithmic}[1]
    			\renewcommand{\algorithmicrequire}{\textbf{Input}:}
			\renewcommand{\algorithmicensure}{\textbf{Output}:}
	
			\REQUIRE The underlying directed graph, \(G(\P,E(\P))\), of the switched system \eqref{e:swsys}, the scalars \(\lambda_{i}\), \(i\in\P\) and \(\mu_{ij}\), \((i,j)\in E(\P)\), the admissible minimum and maximum dwell times on the subsystems, \(\delta\) and \(\Delta\).
			\ENSURE A contractive cycle, \(W\) on \(G\), if exists.

            \STATE Associate edge weights, \(\alpha:E(\P)\to\R\), to \(G\), as follows:
            \begin{align*}
                \alpha(i,j) =
                \begin{cases}
                    \ln\mu_{ij} - \abs{\ln\lambda_{i}}\Delta,\:\:\text{if}\:i\in\PS,\\
                    \ln\mu_{ij} + \abs{\ln\lambda_{i}}\delta,\:\:\text{if}\:i\in\PU.
                \end{cases}
            \end{align*}

            \STATE Apply a negative cycle detection algorithm to \(G\) to obtain a cycle \(W = v_{0},(v_{0},v_{1}),v_{1},\ldots,v_{\ell-1},(v_{\ell-1},v_{0}),v_{0}\) that satisfies \(\displaystyle{\sum_{\substack{{k=0}\\{v_{\ell}:=v_{0}}}}^{\ell-1}\alpha(v_{k},v_{k+1})<0}\).
            \end{algorithmic}
   \end{algorithm}

\subsection{Data-based design of multiple Lyapunov-like functions}
\label{ss:mlf_design}
    Notice that the existence of contractive cycles on the underlying directed graph of the switched system \eqref{e:swsys} depends on three factors: (i) the scalars \(\lambda_{i}\), \(i\in\P\) and \(\mu_{ij}\), \((i,j)\in E(\P)\), (ii) the connectivity of \(G\), and (iii) the scalars, \(\delta\) and \(\Delta\). While (ii) and (iii) are pre-specified, we require a mechanism to design (i) by employing the available subsystems data. The remainder of this section caters to the above purpose.

    Let \(x_{i}(T)\) denote the \(T\)-th element of the trace \(x(0),x(1),\ldots,x(L)\) for subsystem \(i\), and \(L+1\in[\delta:\Delta]\). We let \(x_{i}^{(p)}(T)\) be the \(p\)-th element of \(x_{i}(T)\), \(p=1,2,\ldots,d\). We define
    \begin{align*}
        q_{i}(T) = \pmat{x_{i}^{(1)}(T+1)\\x_{i}^{(1)}(T)\\\vdots\\x_{i}^{(d)}(T)},\:\:T=0,1,\ldots,L-1.
    \end{align*}
    Let
    \begin{align*}
        \Psi_{i} = \pmat{q_{i}(T) & q_{i}(T+1) & \cdots & q_{i}(T+d-1)}\in\R^{(d+1)\times d},
    \end{align*}
    \(T\in\{0,1,\ldots,L-1\}\) be such that the column vectors are linearly independent.

    Fix \(i\in\P\). Notice that whether \(\Psi_{i}\) is well-defined or not, depends on the initial value, \(x_{i}(0)\), and the length, \(L+1\), of the available trace \((x_{i}(0),x_{i}(1),\ldots,x_{i}(L))\). Indeed, to design \(\Psi_{i}\), we need \(L\geq d\) and \(x_{i}(0)\) is such that the vectors \(q_{i}(T),q_{i}(T+1),\ldots,q_{i}(T+d-1)\) are defined and linearly independent for some \(T\in\{0,1,\ldots,L-1\}\). From \cite[Lemma 3]{Park2004} it follows that for every \(A_{i}\) in the companion form described in \eqref{e:comp_form}, there exists \(x_{i}(0)\in\R^{d}\) such that \(\Psi_{i}\) is well-defined with \(T=0\) and \(L=d\). We will, therefore, operate under
    \begin{assump}
    \label{assump:max_dwell_time}
        \(\Delta\geq d+1\).
    \end{assump}

    \begin{lemma}
    \label{lem:auxres2}
    \rm{
        A subsystem \(i\in\PS\) if and only if there exists a symmetric and positive definite matrix \(P_{i}\in\R^{d\times d}\) and a scalar \(0<\lambda_{i}<1\) such that
        \begin{align}
        \label{e:key_ineq}
            \Psi_{i}^\top\pmat{I_{n} & 0_{n}\\0_{n} & I_{n}}^\top\pmat{P_{i} & \overline{0}_{n}\\\overline{0}_{n} & -\lambda_{i}P_{i}}\pmat{I_{n} & 0_{n}\\0_{n} & I_{n}}\Psi_{i} \prec 0.
        \end{align}
    }
    \end{lemma}

    {\bf Proof}:
        It follows from \cite[Theorem 2]{Park2009} that condition \eqref{e:Lyap_eq2} with \(i\in\P_S\) is equivalent to \eqref{e:key_ineq}.

    \begin{lemma}
    \label{lem:auxres3}
    \rm{
        A subsystem \(i\in\PU\) if and only if there exists a symmetric and positive definite matrix \(P_{i}\in\R^{d\times d}\) and a scalar \(\lambda_{i}\geq 1\) such that \eqref{e:key_ineq} holds.
    }
    \end{lemma}

    {\bf Proof}:
        Fix \(i\in\PU\). We will show that condition \eqref{e:Lyap_eq2} is equivalent to \eqref{e:key_ineq}. There exists \(0 < \eta_{i} < 1\) such that \(\eta_{i}A_{i}\) is Schur stable.

        Now, condition \eqref{e:Lyap_eq2} with \(\gamma_{i}(t+1) = \eta_{i}A_{i}\gamma_{i}(t)\) and \(\lambda_{i} = \overline{\lambda}_{i}\) can be rewritten as
        \[
            \gamma_{i}(t)^\top \eta_{i}A_{i}^\top P_{i}\eta_{i}A_{i}\gamma_{i}(t)\leq\overline{\lambda}_{i}\gamma_{i}(t)^\top P_{i}\gamma_{i}(t),
        \]
        or equivalently,
        \[
            \gamma_{i}(t)^\top A_{i}^\top P_{i}A_{i}\gamma_{i}(t)\leq\frac{\overline{\lambda}_{i}}{\eta_{i}^{2}}\gamma_{i}(t)^top P_{i}\gamma_{i}(t).
        \]

        In view of Lemma \ref{lem:auxres2}, the satisfaction of the above inequality is equivalent to the existence of \(P_{i}\succ 0\) and \(\displaystyle{\frac{\overline{\lambda}_{i}}{\eta_{i}^{2}}}\in\R\) such that condition \eqref{e:Lyap_eq2} holds with \(\displaystyle{\lambda_{i} = \frac{\overline{\lambda}_{i}}{\eta_{i}^{2}}}\). Since \(0<\overline{\lambda}_{i},\eta_{i}<1\), it follows that \(\lambda_{i}\geq 1\).

    Lemmas \ref{lem:auxres2} and \ref{lem:auxres3} provide a mechanism to design pairs \((\lambda_{i},P_{i})\), \(i\in\P\), defined in Fact \ref{fact:key1}, from the available subsystems data. The scalars \(\mu_{ij}\), \((i,j)\in E(\P)\) can then be computed by employing Proposition \ref{prop:mu_estimate}.

    We are now in a position to present our solution to Problem \ref{prob:mainprob}.
\section{Result}
\label{s:mainres}
    Given the set of admissible switches between the subsystems, \(E(\P)\), the admissible minimum and maximum dwell times on the subsystems, \(\delta\) and \(\Delta\), and a finite traces of state trajectories for each subsystem, Algorithm \ref{algo:main_algo} designs a stabilizing periodic switching logic \(\sigma\in\Sw\) by employing the following steps:
    \begin{itemize}[label=\(\circ\),leftmargin = *]
        \item First, the underlying directed graph, \(G(\P,E(\P))\), of the switched system \eqref{e:swsys} is constructed from the knowledge of \(\P\) and \(E(\P)\).\footnote{Notice that \(\P\) can be constructed from the knowledge of \(E(\P)\). Indeed, if there is a subsystem \(i\in\P\) such that there is no \(j\in\P\), \(j\neq p\) with \((j,i)\in E(\P)\), then a well-defined switching logic does not activate \(i\) anyway.}
        \item Second, the matrices \(\Psi_{i}\), \(i\in\P\) are constructed by employing the available subsystems data.
        \item Third, the matrices \(\Psi_{i}\), \(i\in\P\) are used to compute the scalars \(\lambda_{i}\), \(i\in\P\) and \(\mu_{ij}\), \((i,j)\in E(\P)\) by employing Lemmas \ref{lem:auxres2}-\ref{lem:auxres3} and Proposition \ref{prop:mu_estimate}, respectively. Notice that the choice of the pairs \((P_{i},\lambda_{i})\), \(i\in\P\) (and hence the corresponding scalars \(\mu_{ij}\), \((i,j)\in E(\P)\)) is not unique. We store all pairs \((P_{i},\lambda_{i})\), \(i\in\P\) that satisfy condition \eqref{e:key_ineq} in \(\Lambda_{i}\), \(i\in\P\), and their corresponding \(\mu_{ij}\), \((i,j)\in E(\P)\) in \(\chi_{ij}\), \((i,j)\in E(\P)\).
        \item Fourth, corresponding to all elements in \(\Lambda_{i}\), \(i\in\P\) and their corresponding elements in \(\chi_{ij}\), \((i,j)\in E(\P)\), Algorithm \ref{algo:contrac_cycle_detection} is applied to \(G\) until a contractive cycle \(W\) is obtained. If \(G\) does not admit any such cycle, then Algorithm \ref{algo:main_algo} reports a failure and terminates. Otherwise, \(W\) is employed to design a stabilizing periodic switching logic that activates the subsystems whose corresponding vertices appear in \(W\) and dwells on each stable and unstable subsystems for \(\Delta\) and \(\delta\) units of time, respectively.
    \end{itemize}
    \begin{algorithm}[htbp]
	\caption{Model-free computation of stabilizing periodic switching logics} \label{algo:main_algo}
    	\begin{algorithmic}[1]
    			\renewcommand{\algorithmicrequire}{\textbf{Input}:}
			\renewcommand{\algorithmicensure}{\textbf{Output}:}
	
			\REQUIRE The sets of indices of the stable and unstable subsystems, \(\PS\) and \(\PU\), the set of admissible switches between the subsystems, \(E(\P)\), the admissible minimum and maximum dwell times on the subsystems, \(\delta\) and \(\Delta\), and a subsystems dataset.
			\ENSURE A stabilizing periodic switching logic, \(\sigma\), or a failure message.

            \STATE {\bf Step I}: Construct the underlying directed graph, \(G(\P,E(\P))\), of the switched system \eqref{e:swsys}.

            \STATE {\bf Step II}: Employ the available subsystems data and construct \(\Psi_{i}\), \(i\in\P\).
            \FOR {\(i=1,2,\ldots,N\)}
                \STATE Construct a well-defined \(\Psi_{i}\).
            \ENDFOR

            \STATE {\bf Step III}: Apply Algorithm \ref{algo:scalar_compute} to compute \(\lambda_{i}\), \(i\in\P\) and \(\mu_{ij}\), \((i,j)\in E(\P)\) from \(\Psi_{i}\), \(i\in\P\).

            \STATE {\bf Step IV}: Detect a contractive cycle on \(G\).
            \FOR {each \((P_{i},\lambda_{i})\in\Lambda_{i}\) and the corresponding \(\mu_{ij}\), \((i,j)\in E(\P)\)}
            \STATE Apply Algorithm \ref{algo:contrac_cycle_detection}.
            \IF {a contractive cycle \(W = v_{0},(v_{0},v_{1}),v_{1},\ldots,v_{\ell-1},(v_{\ell-1},v_{0}),v_{0}\) on \(G\) is detected}
            \STATE Go to Step V.
            \ELSE
            \STATE Output FAIL and terminate.
            \ENDIF
            \ENDFOR

            \STATE {\bf Step V}: Design a stabilizing periodic switching logic.
            \STATE Apply Algorithm \ref{algo:swsig_construc} with \(W\) obtained in Step VI, and obtain a \(\sigma\) with \(\sigma(\tau_{k}) = \Delta\), if \(\sigma(\tau_{k})\in\PS\) and \(\sigma(\tau_{k}) = \delta\), if \(\sigma(\tau_{k})\in\PU\).
    \end{algorithmic}
    \end{algorithm}
	
    Observe that solving \eqref{e:key_ineq} with both \(P_{i}\) and \(\lambda_{i}\), \(i\in\P\), unknown is a numerically difficult task. To address this issue, in Algorithm \ref{algo:scalar_compute}, we employ a line search technique \cite{LMI_book} as follows:
   \begin{itemize}[label=\(\circ\),leftmargin = *]
        \item For \(i\in\PS\), a finite set of values of \(\lambda_{i}\) on the interval \(]0,1[\) is fixed, and corresponding to each element of this set, the feasibility problem \eqref{e:feasprob} is solved for symmetric and positive definite matrices, \(P_{i}\), \(i\in\P\).
        \item For \(i\in\PU\), a finite set of values of \(\eta_{i}\) on the interval \(]0,1[\) is fixed, and corresponding to each element of this set, Schur stability of the matrices, \(\eta_{i}A_{i}\), is checked. The favourable values of \(\eta_{i}\) are stored and the feasibility problem \eqref{e:feasprob} is solved for symmetric and positive definite matrices, \(P_{i}\), \(i\in\P\) with \(\lambda_{i} = \frac{1}{\eta_{i}^{2}}\).
    \end{itemize}
    From Lemma \ref{lem:stab}, it follows that a switching logic, \(\sigma\), obtained from Algorithm \ref{algo:main_algo} obeys the given restrictions on the admissible switches between the subsystems and the admissible dwell times on the subsystems, and preserves GAS of the switched system \eqref{e:swsys}.
    \begin{algorithm}[htbp]
	\caption{Computation of \(\lambda_{i}\), \(i\in\P\) and \(\mu_{ij}\), \((i,j)\in E(\P)\) from \(\Psi_{i}\), \(i\in\P\)} \label{algo:scalar_compute}
    	\begin{algorithmic}[1]
    			\renewcommand{\algorithmicrequire}{\textbf{Input}:}
			\renewcommand{\algorithmicensure}{\textbf{Output}:}
	
			\REQUIRE The matrices \(\Psi_{i}\), \(i\in\P\).
			\ENSURE Sets of scalars \(\lambda_{i}\), \(i\in\P\) and \(\mu_{ij}\), \((i,j)\in E(\P)\).

            \STATE {\bf Step I}: Compute \((P_{i},\lambda_{i})\) from \(\Psi_{i}\), \(i\in\PS\).
            \STATE Fix \(h_{s}>0\) (small enough) and compute \(k_{s}\in\N\) such that \(k_{s}\) is the largest integer satisfying \(k_{s}h_{s}<1\).
            \FOR {all \(i\in\PS\)}
                \STATE Set \(\Lambda_{i} = \emptyset\).
                \FOR {\(\lambda_{i} = h_{s},2h_{s},\ldots,k_{s}h_{s}\)}
                    \STATE Solve the following feasibility problem in \(P_{i}\):
                    \begin{align}
	           		\label{e:feasprob}
		                  \minimize\:\:&\:\:1\nonumber\\
		              	  \sbjto\:\:&\:\:
		                  \begin{cases}
 			                \text{condition}\:\eqref{e:key_ineq},\\
                                        P_{i}^\top = P_{i} \succ 0.\\
		                  			\end{cases}
	             	\end{align}
                    \IF {a solution to \eqref{e:feasprob} exists,}
                        \STATE Set \(\Lambda_{i} = \Lambda_{i}\cup\{(P_{i},\lambda_{i})\}\).
                    \ENDIF
                \ENDFOR
            \ENDFOR

            \STATE {\bf Step II}: Compute \((P_{i},\lambda_{i})\) from \(\Psi_{i}\), \(i\in\PU\).
            \STATE Fix \(h_{u} > 0\) (small enough) and compute \(k_{u}\in\N\) such that \(k_{u}\) is the largest integer satisfying \(k_{u}h_{u} < 1\).
            \FOR {all \(i\in\PU\)}
            \STATE Set \(\Gamma_{i} = \emptyset\) and \(\Lambda_{i} = \emptyset\).
            \FOR {\(\eta_{i} = h_{u},2h_{u},\ldots,k_{u}h_{u}\)}
            \IF {\(\eta_{i}A_{i}\) is Schur stable}
                \STATE Set \(\Gamma_{i} = \Gamma_{i}\cup\{\frac{1}{\eta_{i}^{2}}\}\).
            \ENDIF
            \ENDFOR
            \FOR {all \(\lambda_{i}\in\Gamma_{i}\)}
            \STATE Solve the feasibility problem \eqref{e:feasprob} in \(P_{i}\).
            \IF {a solution to \eqref{e:feasprob} exists}
            \STATE Set \(\Lambda_{i} = \Lambda_{i}\cup\{(P_{i},\lambda_{i})\}\).
            \ENDIF
            \ENDFOR
            \ENDFOR
                				
            \STATE {\bf Step III}: Compute \(\mu_{ij}\), \((i,j)\in E(\P)\) from \(P_{i}\), \(i\in\P\).
            \STATE Set \(\chi_{ij} = \emptyset\).
            \FOR {all \((i,j)\in E(\P)\)}
            \FOR {each pair \((P_{i},\lambda_{i}),(P_{j},\lambda_{j})\in\Lambda_{i}\)}
            \STATE Compute \(\mu_{ij} = \lambda_{\max}(P_{j}P_{i}^{-1})\).
            \STATE Set \(\chi_{ij} = \chi_{ij}\cup\{\mu_{ij}\}\).
            \ENDFOR
            \ENDFOR
    \end{algorithmic}
    \end{algorithm}

    It is worth highlighting that only finitely many possibilities of the pairs \((P_{i},\lambda_{i})\), \(i\in\P\) and their corresponding \(\mu_{ij}\), \((i,j)\in E(\P)\) are explored in Algorithm \ref{algo:main_algo}. Consequently, a failure message obtained from Algorithm \ref{algo:main_algo} does not indicate the non-existence of a set of pairs \((P_{i},\lambda_{i})\), \(i\in\P\), such that \(G\) admits a contractive cycle. In addition, Lemma \ref{lem:stab} provides a sufficient condition for the stability of \eqref{e:swsys}. As a result, a failure message obtained from Algorithm \ref{algo:main_algo} also does not conclude the non-existence of stabilizing elements in \(\Sw\).
\section{A numerical example}
\label{s:numex}
    Consider a switched system \eqref{e:swsys} with \(E(\P) = \{(1,2),(1,4),(1,5)\),\((2,3)\),\((2,4),(2,5),(3,4),(3,5),(4,5)\),\((5,1),(5,4)\}\), \(\delta = 2\) and \(\Delta = 6\). The numerical values of \(A_{i}\), \(i\in\P = \{1,2,3,4,5\}\), are given in Table \ref{tab:data_set1}; they are generated by choosing elements from the interval \([-1,1]\) uniformly at random.
    \begin{table*}[htbp]
	\centering
    \scalebox{1}{
	\begin{tabular}{|c | c | c|}
		\hline
		\(i\) & \(A_{i}\)\\
		\hline
		\(1\) & \(\pmat{ 0.6800919 & -0.0450156 &  0.8332522 &  0.2474328 &  0.6490691\\
                    1 & 0 & 0 & 0 & 0\\
                    0 & 1 & 0 & 0 & 0\\
                    0 & 0 & 1 & 0 & 0\\
                    0 & 0 & 0 & 1 & 0}\)\\

		\hline
        \(2\) & \(\pmat{-0.5271101  & 0.3123521 & -0.6457525 &  0.8071877 &  0.6592868\\
                    1 & 0 & 0 & 0 & 0\\
                    0 & 1 & 0 & 0 & 0\\
                    0 & 0 & 1 & 0 & 0\\
                    0 & 0 & 0 & 1 & 0}\)\\
        \hline
        \(3\) & \(\pmat{-0.442885  & 0.5338433 &  0.6484834  & 0.8689565 &  0.3665952\\
                    1 & 0 & 0 & 0 & 0\\
                    0 & 1 & 0 & 0 & 0\\
                    0 & 0 & 1 & 0 & 0\\
                    0 & 0 & 0 & 1 & 0}\)\\
        \hline
        \(4\) & \(\pmat{0.6482512 & 0.0272578 & -0.4435161 & 0.1849962 & 0.053028\\
                    1 & 0 & 0 & 0 & 0\\
                    0 & 1 & 0 & 0 & 0\\
                    0 & 0 & 1 & 0 & 0\\
                    0 & 0 & 0 & 1 & 0}\)\\
        \hline
        \(5\) & \(\pmat{0.2486157 &  0.0809103 & -0.0931076 &  0.3252463 & -0.1238403\\
                    1 & 0 & 0 & 0 & 0\\
                    0 & 1 & 0 & 0 & 0\\
                    0 & 0 & 1 & 0 & 0\\
                    0 & 0 & 0 & 1 & 0}\)\\
        \hline
	\end{tabular}}
	\caption{Description of the subsystems.}\label{tab:data_set1}
	\end{table*}

\begin{figure}
   \centering
        \includegraphics[scale = 0.4]{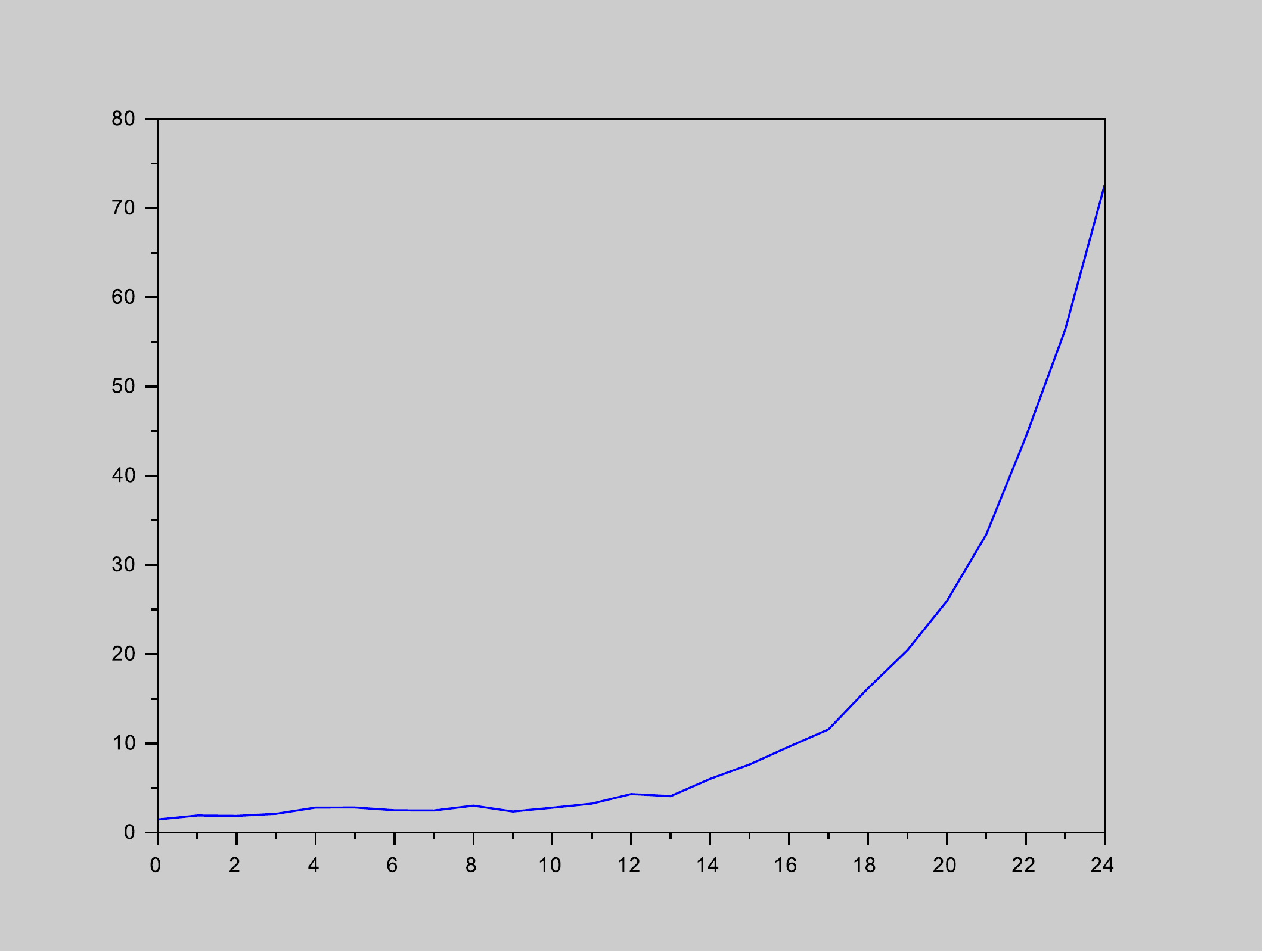}
        \caption{Plot of \((\norm{x(t)})_{t\in\N_{0}}\) under a randomly designed \(\sigma\in\Sw\).}\label{fig:x1_plot}
   \end{figure}

    We first demonstrate that the switched system under consideration is not stable under all \(\sigma\in\Sw\). Towards this end, we fix \(\tau_{0} = 0\), pick \((\sigma(\tau_{k}),\sigma(\tau_{k+1}))\in E(\P)\), \(\tau_{k+1}-\tau_{k}\in[\delta:\Delta]\), \(k=0,1,2,\ldots\), uniformly at random, and plot \((\norm{x(t)})_{t\in\N_{0}}\) corresponding to \(x(0)\in\R^{5}\) chosen uniformly at random from the interval \([-1,1]^{5}\) and the \(\sigma\) designed above. We observe instability of \eqref{e:swsys}, see Figure \ref{fig:x1_plot}. It is, therefore, of interest to identify stabilizing elements of \(\sigma\).

    The values of \(a_{i,j}\), \(j=0,1,\ldots,d-1\), \(i\in\P\) are not known, but the set \(E(\P)\) and the scalars \(\delta\), \(\Delta\) are known. In addition, we have access to subsystems dataset. We will apply Algorithm \ref{algo:main_algo} to design a stabilizing periodic switching logic \(\sigma\in\Sw\). It involves the following steps:

    \begin{enumerate}[label = Step \Roman*., leftmargin = *]
        \item The underlying directed graph, \(G(\P,E(\P))\), of the switched system \eqref{e:swsys} is constructed with \(\P\) and \(E(\P)\) as described above.
        \item The matrices, \(\Psi_{i}\), \(i\in\P\), are constructed by employing the available dataset. The numerical values of \(\Psi_{i}\), \(i\in\P\) are given in Table \ref{tab:data_set2}.
        \item Algorithm \ref{algo:scalar_compute} is applied with \(h_{s} = h_{u} = 0.1\) to compute the sets \(\Lambda_{i}\), \(i\in\P\) and \(\chi_{ij}\), \((i,j)\in E(\P)\).\footnote{The feasibility problem \eqref{e:feasprob} is solved using the lmisolver tool in Scilab 6.0.2.}
        \item Algorithm \ref{algo:contrac_cycle_detection} is applied on \(G\). A contractive cycle \(W = 4, (4,5), 5, (5,4), 4\) is obtained with \(\lambda_{4}=\lambda_{5} = 0.7\),
            \begin{align*}
            {{P_{4} = \pmat{{3481.4063} & {-1322.4505} & {-349.63603} &  {805.92625} &  {5.8206481}\\
                        {-1322.4505} &  {2144.6438} & {-369.98414} & {-465.86262} & {186.60695}\\
                        {-349.63603} & {-369.98414} & {999.52474} & {-208.06702} & {-135.54574}\\
                        {805.92625} &  {-465.86262} & {-208.06702} &  {594.47594} & {-20.831212}\\
                        {5.8206481} &  {186.60695} & {-135.54574} & {-20.831212} &  {140.49607}}}},\\
            {{P_{5} = \pmat{{3521.175} &  {-60.946413} & {-304.62761} &  {368.42809} & {-457.01911}\\
                            {-60.946413} &  {1337.4811} & {-130.53998} & {-179.42523} &  {0.623482}\\
                            {-304.62761} & {-130.53998} &  {779.40748} & {-138.61513} & {-35.892171}\\
                            {368.42809} & {-179.42523} & {-138.61513} &  {489.07866} & {-167.19363}\\
                            {-457.01911} &  {0.623482} &  {-35.892171} & {-167.19363} &  {189.56074}}}},
            \end{align*}
            \(\mu_{45} = 3.1883122\) and \(\mu_{54} = 3.60176\). A switching logic \(\sigma\) is constructed by employing Algorithm \ref{algo:swsig_construc}. \(\sigma\) activates subsystems \(4\) and \(5\) alternatively, and during each activation dwell on them for \(6\) units of time.
    \end{enumerate}

    \begin{table*}[htbp]
	\centering
    \scalebox{1}{
	\begin{tabular}{|c | c | c|}
		\hline
		\(i\) & \(\Psi_{i}\)\\
		\hline
		\(1\) &
            \(\pmat{ 1.1109433 &  0.1262875 &  1.1877076  & 1.7062463 &  1.9976929\\
   0.7867279 &  1.1109433 &  0.1262875 &  1.1877076 &  1.7062463\\
  -0.3330575 &  0.7867279 &  1.1109433 &  0.1262875 &  1.1877076\\
   0.8915796 & -0.3330575 &  0.7867279 &  1.1109433 &  0.1262875\\
  -0.8272248 &  0.8915796 & -0.3330575 &  0.7867279 &  1.1109433\\
   0.0349348 & -0.8272248 &  0.8915796 & -0.3330575 &  0.7867279}\)\\
        \hline
        \(2\)  &
                    \(\pmat{0.053192 &  -1.4739238 &  0.5816219 & -0.0585455 &  1.4498912\\
   0.3680257  & 0.053192 &  -1.4739238 &  0.5816219 & -0.0585455\\
   0.6760328  & 0.3680257  & 0.053192 &  -1.4739238 &  0.5816219\\
  -0.7886473 &  0.6760328 &  0.3680257 &  0.053192 &  -1.4739238\\
  -0.7397434 & -0.7886473 &  0.6760328 &  0.3680257  & 0.053192\\
   0.1878737 & -0.7397434 & -0.7886473 &  0.6760328 &  0.3680257}\)\\
        \hline
        \(3\) &
                \(\pmat{ 0.6482342 &  0.459006  &  0.9236082 &  0.5121273 &  1.1249577\\
  -0.0060993 &  0.6482342 &  0.459006  &  0.9236082 &  0.5121273\\
   0.7121564 & -0.0060993 &  0.6482342 &  0.459006  &  0.9236082\\
   0.4527117 &  0.7121564 & -0.0060993 &  0.6482342 &  0.459006\\
  -0.2887465 &  0.4527117 &  0.7121564 & -0.0060993 &  0.6482342\\
   0.6074417 & -0.2887465 &  0.4527117 &  0.7121564 & -0.0060993}\)\\
        \hline
        \(4\)  &
                \(\pmat{-0.4584539 & -0.0066052 &  0.1112462 &  0.1332211 & -0.0226489\\
                -0.5687414 & -0.4584539 & -0.0066052 &  0.1112462 &  0.1332211\\
                -0.6945576 & -0.5687414 & -0.4584539 & -0.0066052 &  0.1112462\\
                0.0805042 & -0.6945576 & -0.5687414 & -0.4584539 & -0.0066052\\
                -0.3177508 &  0.0805042 & -0.6945576 & -0.5687414 & -0.4584539\\
                0.4460485 & -0.3177508 &  0.0805042 & -0.6945576 & -0.5687414}\)\\
        \hline
        \(5\)  &
                \(\pmat{0.0308642 & -0.1437207 & -0.0167672 & -0.0429631 &  0.0304562\\
                    -0.1540222 &  0.0308642 & -0.1437207 & -0.0167672 & -0.0429631\\
                    -0.2083555 & -0.1540222 &  0.0308642 & -0.1437207 & -0.0167672\\
                    -0.56438  &  -0.2083555 & -0.1540222 &  0.0308642 & -0.1437207\\
                    -0.2037382 & -0.56438  &  -0.2083555 & -0.1540222 &  0.0308642\\
                    -0.8053248 & -0.2037382 & -0.56438 &  -0.2083555 & -0.1540222}\)\\
        \hline
	\end{tabular}}
	\caption{Description of \(\Psi_{i}\), \(i\in\P\).}\label{tab:data_set2}
	\end{table*}

\begin{figure}
   \centering
        \includegraphics[scale = 0.4]{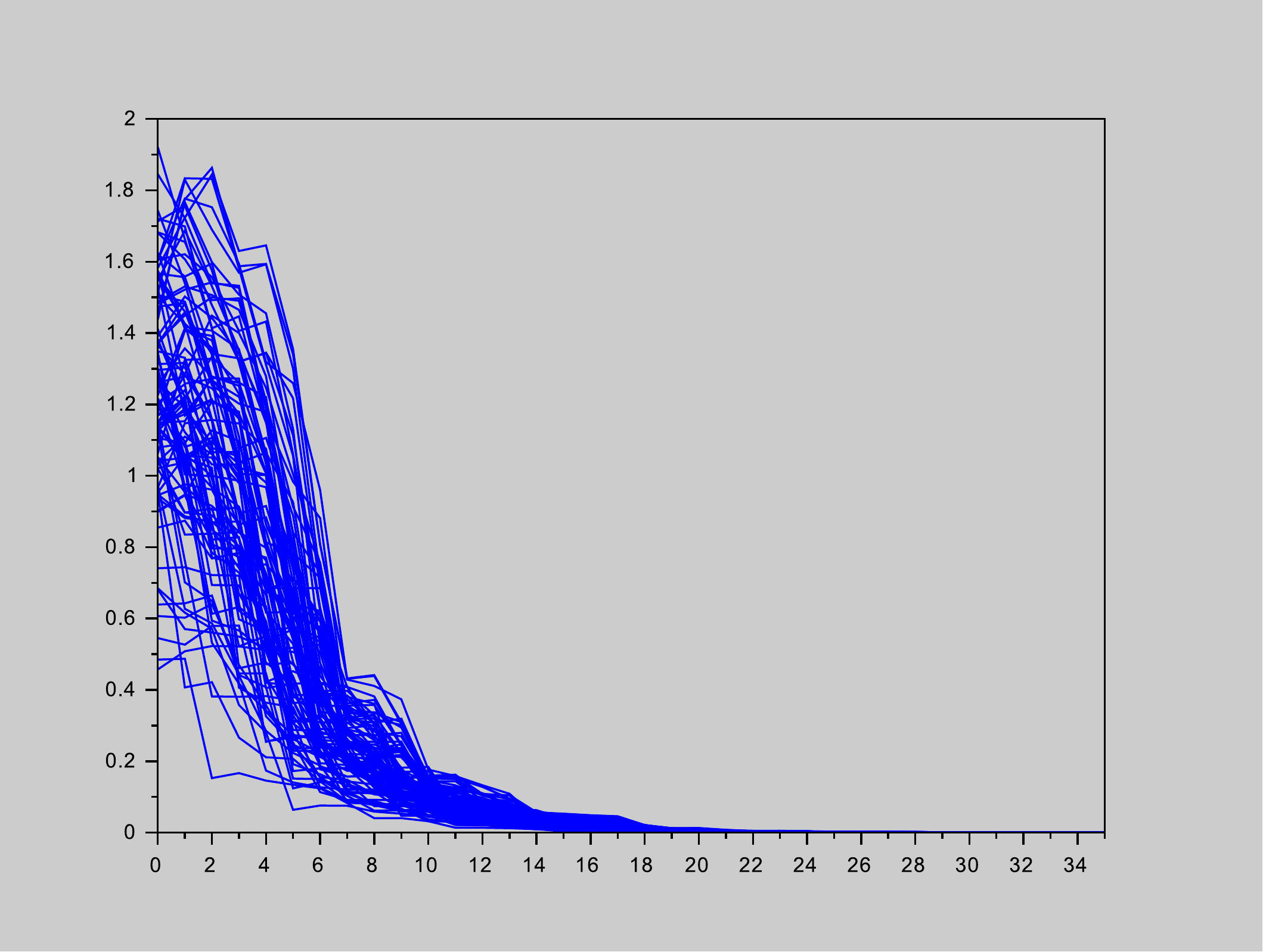}
        \caption{Plot of \((\norm{x(t)})_{t\in\N_{0}}\) under \(\sigma\) obtained from Algorithm \ref{algo:main_algo}.}\label{fig:x_plot}
   \end{figure}

    We now pick \(100\) different \(x(0)\in\R^{5}\) from the interval \([-1,1]^{5}\) uniformly at random, and plot \(\norm{x(t)}_{t\in\N_{0}}\) under the switching logic \(\sigma\) obtained from Algorithm \ref{algo:main_algo}, see in Figure \ref{fig:x_plot}. GAS of the switched system \eqref{e:swsys} is demonstrated.
\section{Conclusion}
\label{s:concln}
    In this paper we presented an algorithm that designs stabilizing periodic switching logics for discrete-time switched linear systems under restricted switching. The proposed design technique can be extended to the setting of stabilizing non-periodic switching logics by employing a cycle \(W = v_{0},(v_{0},v_{1}),v_{1},\ldots,v_{\ell-1},(v_{\ell-1},v_{0}),v_{0}\) that is contractive on \(G\) with multiple choices of \(D_{v_{k}}\in[\delta:\Delta]\), \(k=0,1,\ldots,\ell-1\) and/or multiple contractive cycles \(W_{1},W_{2},\ldots,W_{p}\) on \(G\), see e.g., \cite[Remarks 6,7,10]{xyz3} for discussions on multiple Lyapunov-like functions and graph walks based design of stabilizing non-periodic switching logics. We are currently investigating data-based design of switching logics when the subsystems structures are not restricted to the companion form described in \eqref{e:comp_form} and the available subsystems data are noisy. The findings will be reported elsewhere.



\end{document}